\documentstyle[sprocl]{article}

\input{psfig}
\def\be{\begin{equation}}
\def\ee{\end{equation}}
\def\bea{\begin{eqnarray}}
\def\eea{\end{eqnarray}}
\begin{document}
\begin{flushright}
 NUB-TH-3171/98\\
CTP-TAMU-03/98\\
\end{flushright} 

\title{Non-universal Soft SUSY Breaking and Dark Matter}

\author{ Pran Nath  }

\address{Department of Physics, Northeastern University, Boston,\\ MA 
02115, USA}

\author{ R. Arnowitt }

\address{Center for Theoretical Physics, Department of Physics,\\
Texas A\&M University, College Station, TX77843, USA}

\maketitle

\abstracts{ 
A brief review of non-universalities in the soft SUSY breaking sector of
supergravity grand unification is given. The effects of these
non-universalities on the analysis of event rates in neutralino nucleus 
scattering are discussed. An analysis is also given of the implications 
of the simultaneous imposition of proton stability 
and dark matter constraints. One finds that these constraints put an upper
limit on the gluino mass. }
  
\section{Introduction}
Supergravity unified models\cite{cham1,applied} depend typically on three
arbitrary functions: the gauge kinetic energy function $f_{\alpha\beta}$,
the Kahler potential, and the superpotential. 
The minimal supergravity model is based  on a flat Kahler potential,
and on a gauge kinetic energy which is independent of the heavy fields.  
These assumptions lead to universal soft supersymmetry breaking 
parameters\cite{cham1,applied}. Under
the constraints of radiative breaking of the electro-weak symmetry one
may choose these parameters to be the universal scalar mass $m_0$, the
universal gaugino mass $m_{1/2}$, the universal trilinear coupling $A_0$,
and tan$\beta=\frac{v_2}{v_1}$, where $v_2$ gives mass to the up quarks,
and $v_1$ gives mass to the down quarks. However, in general the Kahler
potential need not be flat, and the presence  of the non-flat Kahler
will lead to non-universalities in the scalar sector\cite{soni,planck}.
Similarly, the universality of the gaugino masses can be broken by the
Planck scale corrections\cite{hill,das}. In the following we will focus
on the non-universalities in the scalar sector. Here
in general one can expand the Kahler potential in terms of the visible
fields so that\cite{soni,planck} 
	\begin{equation}
	K=\kappa^2K_0+K^a_b Q_aQ^b+(K^{ab}Q_a Q_b+ h.c.)+..
	\end{equation}
	where  $Q_a$ are the fields of the visible sector, $Q^a=Q_a^{\dagger}$,
	 and the 
	quantities  $K_0, K_b^a, ..$ etc depend on the fields
	in the hidden sector. As mentioned already in minimal supergravity 
	unification one makes 
	the assumption of a flat Kahler, i.e., 
	 $K^a_b=K(h,h^{\dagger})\delta^a_b$ 
	 (where h are the hidden sector fields) and vanishing 
	 non-holomorphic terms beyond the quadratic order. This assumption then
	 leads to universal soft SUSY breaking at the SUSY breaking scale.
	 However,  due to dynamics at the Planck scale
	 one will have in general a non-flat Kahler potential
	 which will lead to non-universalities. There are, 
	 however, rather stringent constraints on non-universalities from
	 flavor changing neutral currents (FCNC).
	 One sector where the non-universalities are not stringently 
	 constrained by FCNC is the Higgs sector. 
	 Here one may parametrize the  non-universalities 
	 at the GUT scale by\cite{matallio,berez,nonuni} 
 \begin{equation}
m_{H_1}^2(M_G) =  m_0^2 (1 + \delta_1),  ~~m_{H_2}^2(M_G) = m_0^2 (1 + \delta_2)
\end{equation}
where a reasonable limit on $\delta_i$ is $|\delta_i|\leq 1$ (i=1,2). 
%It turns out, however, that the non-universalities in the Higgs sector
%and in the third generation sector are highly coupled and the
%non-universalities in the third generation sector should be considered 
%along with the non-universalities in the Higgs sector.  The 
%non-universalities in the third generation sector can be parametrized in 
%a manner analogous to the non-universalities in the Higgs sector, i.e.,
However,
it was pointed out in ref.\cite{nonuni} that the non-universalities in
the Higgs sector and in the third generation sector are strongly coupled.
Thus one should also include non-universalities in the third generation
sector 
in a manner analogous to the non-universalities in the Higgs sector, i.e.,
 $ m_{\tilde Q_L}^2=m_0^2(1+\delta_3), ~~m_{\tilde U_R}^2=m_0^2(1+\delta_4)$,
%\end{equation}  
where again  a reasonble limit on the $\delta_3, \delta_4$ is 
$|\delta_i|\leq 1$(i=3,4). 

 Non-universalities affect low energy physics in important ways. One
 of the  parameters which controls low energy physics is the Higgs 
 mixing parameter $\mu$.
 One can analytically display the effects of non-universalities on 
 $\mu^2$ for the case of small tan$\beta$ when the effects of b quark 
 couplings can be ignored. One finds in this case that the non-universality
 correction to $\mu$  can be written in the form\cite{nonuni}

\begin{equation}
\Delta\mu^2=m_0^2
\frac{1}{t^2-1}(\delta_1-\delta_2t^2-\frac{ D_0-1}{2}(\delta_2 +
\delta_3+\delta_4)t^2)+ \frac{3}{5}\frac{t^2+1}{t^2-1}S_0p
\end{equation}
Here 
$D_0=(1-(\frac{m_t}{m_f})^2), ~~m_f\simeq 200 sin\beta ~GeV$, 
 t$\equiv tan\beta$, 
$S_0=Tr(Ym^2)$, and p=0.0446.
The Tr($Ym^2$) term is the anomaly term which vanishes for the 
universal case since Tr(Y)=0, but more generally it makes  a 
contribution although this contribution is relatively small.

\section{Analysis of Event Rates in Dark Matter Detectors}
One of the interesting  features of supersymmetric models is that 
with R parity invariance they produce a lowest supersymmetric particle 
which is absolutely stable and hence a candidate for cold dark matter (CDM).
Further in supergravity unification one finds that over most of the
parameter space of the model the lightest neutralino is also the LSP.
The  neutralino is  in general a combination of two neutral gaugino
states and two neutral Higgsino states, i.e., one has\cite{jungman}
\begin{equation}
\chi_1^0=n_{11}\tilde W_3+ n_{12}\tilde B+ n_{13}\tilde H_1+n_{14}\tilde H_2
\end{equation}
where $\tilde W_3$ is the neutral component of the SU(2) gauginos, 
$\tilde B$ is the U(1) gaugino, and  $\tilde H_i$ (i=1,2) 
are the neutral Higgsinos.
The analysis of relic density has been discussed extensively in the 
literature. For some recent references see\cite{lopez,greist,accurate}.

\begin{figure}
%\rule{5cm}{0.2mm}\hfill\rule{5cm}{0.2mm}
\vskip -1.5cm
%\rule{5cm}{0.2mm}\hfill\rule{5cm}{0.2mm}
\psfig{figure=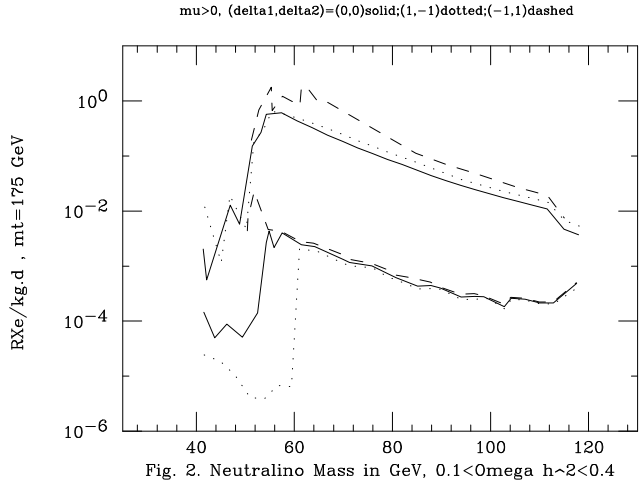,height=2.5in}
\caption{Maximum and minimum of event rates/kg.da for xenon.
The analysis is for the cases (i)$\delta_1=0=\delta_2$ (solid),
(ii)$\delta_1=1=-\delta_2$ (dotted), and 
(iii) $\delta_1=-1=-\delta_2$ (dashed), when $\mu>0$, 
$0.1<\Omega h^2<0.4$, and $m_t=175$ GeV. (Taken from ref.[9]). 
\label{fig:radis} }
\end{figure}

 	One can put reasonably stringent bounds on the relic density.
 	Using the current astro-physical data one finds that the 
 	relic density is constrained by the following 
 	limits 
 	\begin{equation}
 	0.1<\Omega h^2<0.4
 	\end{equation}
 	There are a variety of techniques discussed in the literature
 	for the detection of supersymmetric dark matter. We shall focus
 	here on the direct method which utilizes the scattering of 
 	CDM from target 
 	nuclei\cite{goodman,bottino,bednyakov,an2,na2,gondolo}.
 	 Event rate analyses for this scattering
 	have been discussed within MSSM and also for the standard 
 	supergravity unified models with four soft SUSY breaking 
 	parameters and including the $b\rightarrow s+\gamma$ 
 	experimental constraint\cite{alam} which has a significant effect
 	on dark matter analyses\cite{bsgamma}. More recently the effect of non-universalities on
 	the event rates has also been investigated. In Fig.1 we 
 	exhibit the result of the analysis for event rates using a xenon
 	target. One finds that in the region of the neutralino mass
 	$<65$ GeV the maximum and the minimum of event rates can 
 	vary by a factor of 10 due to the effect of non-universalities.
 	However, in the region of the neutralino mass $> 65$ GeV 
 	the effect of  the non-universalities
 	on the maximum and the minimum event rate curves is relatively
 	smaller. This phenomenon occurs in part due to the relatively
 	 large contribution that the Landau pole arising from the top
 	 Yukawa coupling makes\cite{landau} in this region. 
 	 The analysis of Fig.1 shows
 	 that the event rates can vary over a wide range from 
 	 few events/kg.d to $10^{-5}$events/kg.d. The current sensitivity
 	 of the detectors is O(1) event/Kg.d\cite{bernabei,bottino}. Thus
 	 one needs detectors more sensitive by a factor of 
 	 $10^3$ or more to sample a majority of the  parameter space of the 
 	 supergravity models\cite{cline}.

 \section{Effects of Proton Stability and Dark Matter Constraints}
	We discuss now the implications of imposing simultaneously
	the constraints of proton stability and relic density. As is
	well known in SUSY unified theories the dominant proton decay 
	proceeds via dimension five operators generated via the exchange
	of color triplet Higgsinos. 
	The  dominant decay mode here is the mode $p\rightarrow \bar \nu K$
	and theoretically one finds that\cite{wein,acn,testing} 
   \begin{equation}
   \Gamma (p\rightarrow \bar\nu K^+)=\sum_{i=e,\mu,\tau}\Gamma(p\rightarrow
   \bar\nu_iK^+)\nonumber\\
   =(\frac{\beta_p}{M_{\tilde H_3}})^2 F  
 %  |A|^2|B|^2C
   \end{equation}
   Here $M_{\tilde H_3}$ is the Higgs triplet mass and  $\beta_p$ is the 
   matrix element of the three quark operator between the vacuum and the
   the proton state. Currently the best evaluation of $\beta_p$
   comes from lattice gauge calculations which gives\cite{gavela}
 $\beta_p=5.6\times 10^{-3} GeV^3$. The term  F in Eq.(6) 
 depends on a variety of factors. These include quark masses and CKM factors,
 SUSY spectrum which enters in dressing loop  diagrams needed to 
 convert dimension five operators into dimension six four fermion 
 operators, and chiral Lagrangian factors needed to convert the four
 fermion operators into an effective Lagrangian involving mesons and  
 baryons. 

\begin{figure}
%\rule{5cm}{0.2mm}\hfill\rule{5cm}{0.2mm}
\vskip -0.5cm
%\rule{5cm}{0.2mm}\hfill\rule{5cm}{0.2mm}
\psfig{figure=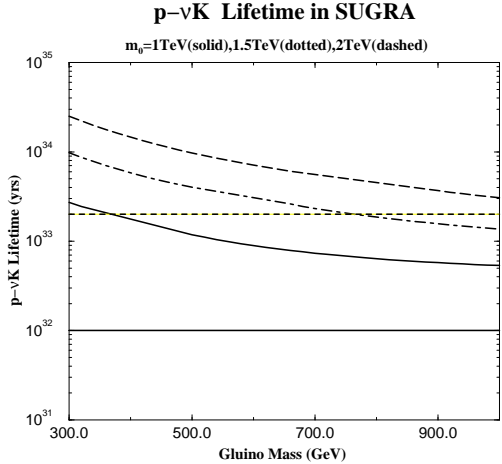,height=2.5in}
\caption{Analysis of the maximum lifetime for the mode 
$p\rightarrow \bar\nu K$ for various choices of the naturalness
assumption on $m_0$, i.e., 1 TeV (solid), 1.5 TeV (dashed-dot),
2.0 TeV (dashed) for the minimal SU(5) model. The lower horizontal
line is the current experimental lower limit and the upper 
horizontal  line is the lower limit expected from Super K.
(Taken from ref.[29]).   
\label{fig:rad}
}
\end{figure}
  
   In Fig.2 we give  an analysis of the maximum lifetime 
   for the decay mode $p\rightarrow \bar \nu K$ for the minimal
   supergravity model as a  function of
   the gluino mass for different choices of naturalness for the 
   parameter $m_0$\cite{gluino}.
    The analysis is done by integrating over the 
   remaining parameters of the model, i.e., $A_0$ and tan$\beta$. 
   The results of the analysis can be compared to the current
   experimental lower limit of\cite{pdg} 
   
   \begin{equation}
   \tau(p\rightarrow \bar\nu K) >1\times 10^{32} yr 
   \end{equation}
 It is expected that Super Kamionkande will reach a 
 sensitivity of\cite{totsuka}   
   \begin{equation}
   \tau(p\rightarrow \bar\nu K)> 2\times 10^{33} yr  
   \end{equation}
   and ICARUS may reach a sensitivity of\cite{icarus} 
   $\tau(p\rightarrow \bar\nu K)> 5\times 10^{33}$ yr. 
  From Fig.2  we find that for the case when the naturalness on $m_0$ is 
   chosen to be 1 TeV  the parameter space of the minimal model for 
   gluino masses $\geq 300$ GeV will be exhausted if Super K 
   can reach their
   expected sensitivity given by Eq.8\cite{gluino}. 
   For the choice of naturalness of
   $m_0=1.5$ TeV, one finds that the parameter space of the minimal
   model for $m_{\tilde g}\geq 750 $ GeV will be exhausted. For the 
   case of the naturalness of $m_0=2$ TeV  one finds that 
  the region of the gluino masses up to 1 TeV is  still allowed.
  For the case of ICARUS the constraints will be even more stringent.   
   
  Next we discuss the effect of imposing the relic density constraint.
  The analysis is given in fig.3. In this case one finds that the gluino 
  mass range consistent with
  proton stabilty and relic density constraint is severely limited.
  For the case when the naturalness on $m_0$ is chosen to be 1 TeV,
  the Kamiokande limit on the gluino mass consistent with p stability 
  and relic 
  density constraint falls significantly below 500 GeV. For the case
  when the naturalness on $m_0$ is increased to 5 TeV, one still finds
  that the gluino mass lies below 500 GeV. The reason for this stringent
  constraint is not difficult to understand. It arises due to the fact that
  for gluino masses greater than 500 GeV, one is in the region beyond
  where the annihilation of neutralinos via the Z boson and the Higgs boson 
  exchange take place. In this region the relic density is governed
  by the sfermion exchange diagrams and an efficient annihilation 
  requires a small value of $m_0$, with typical vaules of 
  $m_0\leq 100$ GeV. However, a small value of $m_0$ tends to destabilize 
  the proton since proton decay has a dependence of the type 
  $\frac{m_{\tilde g}}{m_0}$. Thus the combined 
  p stability and relic density constraints  appear to put a stringent
  limit on the gluino mass. In this sense the minimal model  in the 
  region of gluino masses $m_{\tilde g}>500$ GeV with the imposition of 
  the relic density constraints is similar to the so called 
  "unflipped no-scale" 
  models which also have problems with p stability\cite{noscale}.\\

   The analysis has been given here for the minimal
  SU(5) supergravity model. However, one expects a similar analysis to
  hold for a class of non-minimal models as well\cite{gluino}.

%% 8888888888888888888888

\begin{figure}
%\rule{5cm}{0.2mm}\hfill\rule{5cm}{0.2mm}
\vskip 0.0cm
%\rule{5cm}{0.2mm}\hfill\rule{5cm}{0.2mm}
\psfig{figure=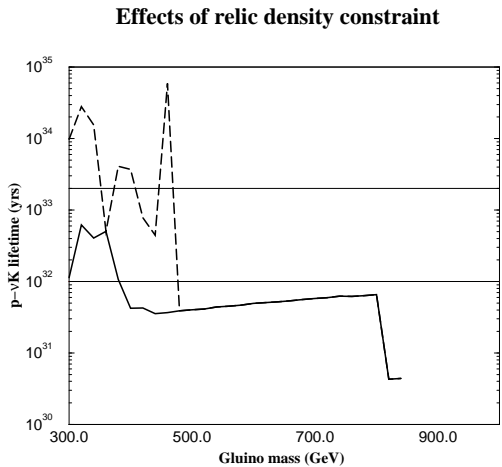,height=2.5in}
\caption{The maximum life of the $p\rightarrow \bar\nu K$ mode
as a function of the gluino mass with the relic density constraint 
$0.1<\Omega h^2 <0.4$ for the naturalness constraint on $m_0$ of 
1 TeV (solid) and 5 TeV (dashed). The horizontal lines  are as in
Fig.2. (Taken from ref.[29]).
\label{fig:radsh}
}
\end{figure}

\section*{Acknowledgements}
 This work was  supported in part by NSF grants PHY-9602074 and
 PHY-9722090.       
        			
%\newpage

\end{document}